\documentclass[showpacs,aps,prc]{revtex4}
\usepackage[T1]{fontenc}
\usepackage[latin1]{inputenc}
\usepackage{graphicx}
\usepackage{amssymb}
\usepackage{verbatim}
\usepackage{epsfig}

\newcommand{\ba}{\begin{eqnarray}}
\newcommand{\ea}{\end{eqnarray}}

\begin{document}
\pagestyle{plain}

\title{Strange form factors of the nucleon
\footnote{Invited talk at XXIX Symposium on Nuclear Physics, 
Hacienda Cocoyoc, Morelos, Mexico, January 3-6, 2006}}  
\author{R. Bijker}
\email{bijker@nucleares.unam.mx}
\affiliation{Instituto de Ciencias Nucleares, 
Universidad Nacional Aut\'onoma de M\'exico, 
AP 70-543, 04510 M\'exico, D.F., M\'exico}

\date{Received January 27, 2006}

\begin{abstract}
Abstract. The strangeness content of nucleon form factors is analyzed in a 
two-component model with a quark-like intrinsic structure surrounded by a 
meson cloud. A comparison with the available experimental data from the 
SAMPLE, PVA4, HAPPEX and G0 collaborations shows a good overall agreement. 


Resumen. 
Se analiza el contenido de extra\~neza de los factores de forma del nucle\'on 
en un modelo de dos componentes que consiste en una estructura 
intr{\'{\i}}nseca de tres cuarks constituyentes rodeada de una nube mes\'onica.  
Una comparaci\'on con los datos experimentales disponibles de las colaboraciones 
SAMPLE, PVA4, HAPPEX y G0 muestra un buen ajuste para los factores de 
forma con extra\~neza.  

Descriptores: Factores de forma extra\~nos, dominancia de mesones vectoriales
\end{abstract}


\maketitle                   

\section{Introduction}

The contribution of the different quark flavors to the electromagnetic 
structure of the nucleon can be studied by combining the nucleon's response 
to the electromagnetic and weak neutral vector currents \cite{Manohar}. 
Especially, the contribution of strange quarks to the nucleon structure 
is of interest because it is exclusively part of the quark-antiquark sea.  

In recent experiments, parity-violating elastic electron-proton 
scattering has been used to probe the contribution of strange quarks 
to the structure of the nucleon \cite{Beise}.  
The strange quark content of the form factors can be determined 
assuming charge symmetry and combining parity-violating asymmetries 
with measurements of the electric and magnetic form factors of the 
proton and neutron. The various experiments are sensitive to different 
combinations of the strange quark contributions to the charge 
distribution and the magnetization represented by the strange electric 
and magnetic form factors. 
There are various methods to disentangle the electric and magnetic 
contributions: by measuring parity-violating asymmetries at both forward 
and backward angles \cite{Spayde}, by using different targets \cite{Aniol06}, 
or by combining parity-violating asymmetries with (anti)neutrino-proton  
scattering data \cite{Pate}. 

The first experimental results from the SAMPLE \cite{Beise,Spayde},  
PVA4 \cite{Maas}, HAPPEX \cite{Aniol06,Aniol04,Aniol05b} and G0 
\cite{Armstrong} collaborations have shown evidence for a nonvanishing 
strange quark contribution to the structure of the nucleon. 
In particular, the strangeness content of the proton magnetic moment was 
found to be positive \cite{Aniol05b}, suggesting that the strange quarks 
reduce the proton's magnetic moment. This is an unexpected and surprising finding, 
since a majority of theoretical studies favors a negative value \cite{beck}. 

The aim of this contribution is to analyze the available experimental 
data on strange form factors in a two-component model of the nucleon.  

\section{Two-component model of nucleon form factors} 
 
Electromagnetic and weak form factors contain the information about the 
distribution of electric charge and magnetization inside the nucleon.  
These form factors arise from matrix elements of the corresponding 
vector current operators
\ba
\left< N \left| V_{\mu} \right| N \right> = \bar{u}_N \left[ 
F_1(Q^2) \, \gamma_{\mu} + \frac{i}{2M_N} F_2(Q^2) \, \sigma_{\mu\nu} q^{\nu} 
\right] u_N ~.
\ea
Here $F_{1}$ and $F_2$ are the Dirac and Pauli form factors 
which are functions of the squared momentum transfer $Q^2=-q^2$. 
The electric and magnetic form factors, $G_{E}$ and $G_{M}$, are 
obtained from $F_{1}$ and $F_{2}$ by the relations $G_E=F_1-\tau F_2$ 
and $G_M=F_1 + F_2$ with $\tau=Q^2/4 M_N^2$. 

Different models of the nucleon correspond to different assumptions about 
the Dirac and Pauli form factors. In this contribution, I consider the 
two-component model of \cite{IJL,BI} in which the external photon 
couples both to an intrinsic three-quark structure described by the form 
factor $g(Q^2)$, and to a meson cloud via vector-meson ($\rho$, $\omega$ 
and $\phi$) dominance (VMD). In the original VMD calculation \cite{IJL}, 
the Dirac form factor was attributed to both the intrinsic structure and 
the meson cloud, and the Pauli form factor entirely to the meson cloud. 
In \cite{BI}, it was shown that the addition of an intrinsic 
part to the isovector Pauli form factor as suggested by studies of 
relativistic constituent quark models in the light-front approach 
\cite{frank}, improves the results for the neutron elecric and magnetic 
form factors considerably. 

In order to incorporate the contribution of the isocalar ($\omega$ and 
$\phi$) and isovector ($\rho$) vector mesons, it is convenient to introduce  
the isoscalar and isovector current operators 
\ba
V_{\mu}^{I=0} &=& \frac{1}{6} \left( \bar{u} \gamma_{\mu} u 
+ \bar{d} \gamma_{\mu} d -2 \bar{s} \gamma_{\mu} s \right) ~,
\nonumber\\ 
V_{\mu}^{I=I} &=& \frac{1}{2} \left( \bar{u} \gamma_{\mu} u 
- \bar{d} \gamma_{\mu} d \right) ~. 
\ea
The corresponding isoscalar Dirac and Pauli form factors depend on 
the couplings to the $\omega$ and $\phi$ mesons 
\ba
F_{1}^{I=0}(Q^{2}) &=& \frac{1}{2} g(Q^{2}) \left[ 
1-\beta_{\omega}-\beta_{\phi} 
+\beta_{\omega} \frac{m_{\omega }^{2}}{m_{\omega }^{2}+Q^{2}} 
+\beta_{\phi} \frac{m_{\phi}^{2}}{m_{\phi }^{2}+Q^{2}}\right] ~, 
\nonumber\\
F_{2}^{I=0}(Q^{2}) &=&\frac{1}{2}g(Q^{2})\left[ 
\alpha_{\omega} \frac{m_{\omega }^{2}}{m_{\omega }^{2}+Q^{2}} 
+ \alpha_{\phi} \frac{m_{\phi}^{2}}{m_{\phi}^{2}+Q^{2}}\right] ~,
\ea
and the isovector ones on the coupling to the $\rho$ meson \cite{BI}
\ba
F_{1}^{I=1}(Q^{2}) &=&\frac{1}{2}g(Q^{2})\left[ 1-\beta_{\rho} 
+\beta_{\rho} \frac{m_{\rho}^{2}}{m_{\rho}^{2}+Q^{2}} \right] ~,  
\nonumber\\
F_{2}^{I=1}(Q^{2}) &=&\frac{1}{2}g(Q^{2})\left[ 
\frac{\mu_{p}-\mu_{n}-1-\alpha_{\rho}}{1+\gamma Q^{2}} 
+\alpha_{\rho} \frac{m_{\rho }^{2}}{m_{\rho}^{2}+Q^{2}} \right] ~.  
\label{ff}
\ea
The proton and neutron form factors correspond to the sum and 
difference of the isoscalar and isovector contributions. 
For the intrinsic form factor a dipole form $g(Q^{2})=(1+\gamma Q^{2})^{-2}$ 
is used whose asymptotic behavior is consistent with p-QCD \cite{pQCD} 
and which coincides with the form used in an algebraic treatment of the 
intrinsic three-quark structure \cite{bijker}. 

The large width of the $\rho$ meson plays an important role for the small 
$Q^{2}$ behavior of the form factors and is taken is taken into account in 
the same way as in \cite{IJL,BI} by the replacement \cite{frazer} 
\begin{equation}
\frac{m_{\rho }^{2}}{m_{\rho }^{2}+Q^{2}}\rightarrow \frac{m_{\rho
}^{2}+8\Gamma _{\rho }m_{\pi }/\pi }{m_{\rho }^{2}+Q^{2}+\left( 4m_{\pi
}^{2}+Q^{2}\right) \alpha(Q^{2}) \Gamma_{\rho }/m_{\pi }}~,  \label{ff3}
\end{equation}
with 
\begin{equation}
\alpha \left( Q^{2}\right) =\frac{2}{\pi }\left[ \frac{4m_{\pi }^{2}+Q^2}{Q^2}%
\right] ^{1/2}\ln \left( \frac{\sqrt{4m_{\pi }^{2}+Q^{2}}+\sqrt{Q^{2}}}{%
2m_{\pi }}\right) ~.  \label{ff4}
\end{equation}
The meson dynamics is important for small values of $Q^2$, whereas 
for large values the form factors satisfy the asymptotic behavior 
of p-QCD, $F_1 \sim 1/Q^4$ and $F_2 \sim 1/Q^6$ \cite{pQCD}.

Since the intrinsic part is associated with the valence quarks of the 
nucleon, the strange quark content of the nucleon form factors arises 
from the meson wave functions 
\ba
\left| \omega \right> &=& \cos \epsilon \left| \omega_0 \right> 
- \sin \epsilon \left| \phi_0 \right> ,
\nonumber\\
\left| \phi \right> &=& \sin \epsilon \left| \omega_0 \right> 
+ \cos \epsilon \left| \phi_0 \right> ,
\ea
where $\left| \omega_0 \right>=\left( u \bar{u} + d \bar{d} \right)/\sqrt{2}$ 
and $\left| \phi_0 \right> = s \bar{s}$ are the ideally mixed states. 
Under the assumption that the strange form factors have the same form as 
the isoscalar ones, the Dirac and Pauli form factors that correspond to the 
strange current 
\ba
V_{\mu}^s = \bar{s} \gamma_{\mu} s ~,
\ea
are expressed as the product  
of an intrinsic part $g(Q^2)$ and a contribution from the meson cloud as 
\ba
F_{1}^{s}(Q^{2}) &=& \frac{1}{2}g(Q^{2})\left[ 
\beta_{\omega}^s \frac{m_{\omega}^{2}}{m_{\omega }^{2}+Q^{2}} 
+\beta_{\phi}^s \frac{m_{\phi}^{2}}{m_{\phi }^{2}+Q^{2}}\right] ~, 
\nonumber\\
F_{2}^{s}(Q^{2}) &=& \frac{1}{2}g(Q^{2})\left[ 
\alpha_{\omega}^s \frac{m_{\omega}^{2}}{m_{\omega }^{2}+Q^{2}}
+\alpha_{\phi}^s \frac{m_{\phi}^{2}}{m_{\phi }^{2}+Q^{2}}\right] ~.
\label{sff}
\ea

The $\beta$'s and $\alpha$'s in Eqs.~(\ref{ff}) and (\ref{sff}) 
are not independent of one another. The coefficients appearing in the 
isoscalar and strange form factors depend on the same nucleon-meson and 
current-meson couplings \cite{Jaffe}. In addition, they are constrained 
by the electric charges and magnetic moments of the nucleon which leads 
to two independent isoscalar couplings
\ba
\alpha_{\omega} &=& \mu_p + \mu_n -1 - \alpha_{\phi} ~,
\nonumber\\
\beta_{\omega} &=& - \beta_{\phi} \tan(\theta_0+\epsilon)/\tan \epsilon ~.
\label{coef1}
\ea
The strange couplings are then given by  
\ba
\beta_{\omega}^s/\beta_{\omega} =  
\alpha_{\omega}^s/\alpha_{\omega} &=& 
-\sqrt{6} \, \sin \epsilon/\sin(\theta_0+\epsilon) ~,
\nonumber\\
\beta_{\phi}^s/\beta_{\phi} =  
\alpha_{\phi}^s/\alpha_{\phi} &=& 
-\sqrt{6} \, \cos \epsilon/\cos(\theta_0+\epsilon) ~.  
\label{coef2}
\ea
where 
$\tan \theta_0 = 1/\sqrt{2}$. The mixing angle $\epsilon$ can be determined from 
the decay properties of the $\omega$ and $\phi$ mesons. Here I take the value 
$\epsilon=0.053$ rad  obtained in \cite{Jain}. 

\section{Results}

In order to calculate the nucleon form factors in the two-component model the 
five coefficients, $\gamma$ from the intrinsic form factor, $\beta_{\phi}$ and 
$\alpha_{\phi}$ from the isoscalar couplings, and $\beta_{\rho}$ and 
$\alpha_{\rho}$ from the isovector couplings, are determined in a least-square 
fit to the electric and magnetic form factors of the proton and the neutron using the 
same data set as in \cite{BI}. The electromagnetic form factor of the proton 
and neutron are found to be in good agreement with experimental data \cite{iguazu}.  
According to Eq.~(\ref{coef2}), the strange couplings can be determined from the 
fitted values of the isoscalar couplings to be 
$\beta_{\phi}^s=-\beta_{\omega}^s=0.202$,  $\alpha_{\phi}^s=0.648$ and 
$\alpha_{\omega}^s=-0.018$ \cite{iguazu}.   

\begin{figure}[htb]
\centerline{\epsfig{file=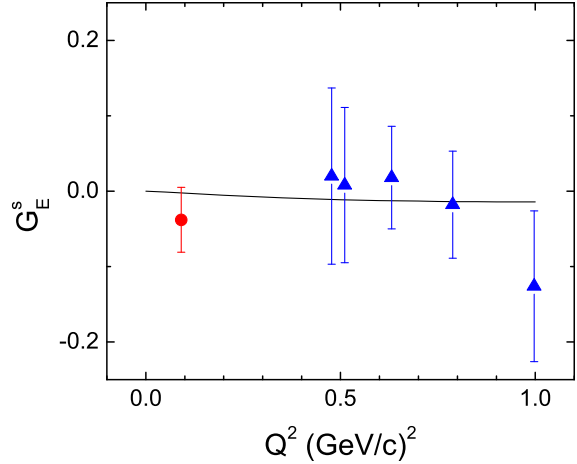,width=0.5\textwidth}} 
\caption[]{\small 
Comparison between theoretical and experimental values of the strange 
electric form factor. The experimental values are taken from \cite{Aniol06} 
(circle) and \cite{Frascati} (triangle).}
\label{GEs}
\end{figure}

\begin{figure}[htb]
\centerline{\epsfig{file=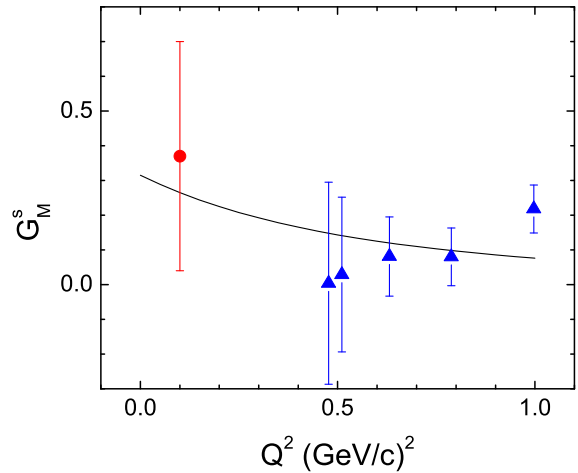,width=0.5\textwidth}} 
\caption[]{\small 
Comparison between theoretical and experimental values of the strange 
magnetic form factor. The experimental values are taken from \cite{Spayde} 
(circle) and \cite{Frascati} (triangle).}
\label{GMs}
\end{figure}

Figs.~\ref{GEs} and \ref{GMs} show the strange electric and magnetic form 
factors as a function of $Q^2$. The theoretical values of 
$G_E^s = F_1^s - \tau F_2^s$ are small and negative.  
The Dirac form factor $F_1^s$ is small due 
to canceling contributions of the $\omega$ and $\phi$ couplings which arose 
as a consequence of the fact that the strange (anti)quarks do not contribute  
to the electric charge $G_E^s(0)=F_1^s(0)=\beta_{\omega}^s+\beta_{\phi}^s=0$.  
Moreover, for this range of momentum transfer the contribution from the Pauli 
form factor $F_2^s$ is suppressed by the factor $\tau=Q^2/4M_N^2$. 
The theoretical values are in good agreement with the recent experimental 
result of the HAPPEX Collaboration in which $G_E^s$ was determined for the 
first time in parity-violating electron scattering from $^{4}$He \cite{Aniol06}. 
The experimental value $G_E^s=-0.038 \pm 0.042 \pm 0.010$ (circle) 
measured at $Q^2=0.091$ (GeV/c)$^2$ is consistent with zero. 

The strange magnetic form factor $G_M^s=F_1^s + F_2^s$ is positive, since 
it is dominated by the contribution from the Pauli form factor. 
The SAMPLE experiment measured the parity-violating asymmetry at backward 
angles, which allowed to determine the strange magnetic form factor at 
$Q^2=0.1$ (GeV/c)$^2$ as $G_M^s=0.37 \pm 0.20 \pm 0.26 \pm 0.07$. 
The other experimental values of $G_E^s$ and $G_M^s$ in Figs.~\ref{GEs} 
and \ref{GMs} were obtained \cite{Pate,Frascati} by combining the 
(anti)neutrino data from E734 \cite{Ahrens} with the parity-violating 
asymmetries from HAPPEX \cite{Aniol04} and G0 \cite{Armstrong}. 
The theoretical values are in good overall agreement with the 
experimental ones for the entire range $0 < Q^2 < 1.0$ (GeV/c)$^2$. 

The strange magnetic moment is calculated to be positive 
\ba
\mu_s = G_M^s(0) = \frac{1}{2} (\alpha_{\omega}^s+\alpha_{\phi}^{s}) 
= 0.315 \, \mu_N ~,
\ea
in units of the nuclear magneton, $\mu_N=e \hbar/2M_N c$. This value is in 
agreement with recent experimental evidence from the SAMPLE collaboration 
\cite{Spayde} and an analysis of the world data $G_M^s=0.55 \pm 0.28$ at  
$Q^2=0.1$ (GeV/c)$^2$ \cite{Aniol05b}. By convention, 
the definition of the strange magnetic moment $\mu_s$ does not involve 
the electric charge of the strange quark.
  
Theoretical calculations of the strange magnetic moment show a large variation, 
although most QCD-inspired models seem to favor a negative value in the range 
$-0.6 \lesssim \mu_s \lesssim 0.0$ $\mu_N$ \cite{beck}. There are relatively 
few calculations that give a sizeable positive strange magnetic moment, ranging 
from $0.074-0.115$ $\mu_N$ in the $SU(3)$ chiral quark soliton model \cite{silva}, 
$0.16 \pm 0.03$ $\mu_N$ in a group theoretical approach with flavor $SU(3)$ 
breaking \cite{jido} to $0.37$ $\mu_N$ in the $SU(3)$ chiral bag model \cite{hong}. 
Recent quenched lattice-QCD calculations give a small value, {\it e.g.}   
$0.05 \pm 0.06$ \cite{lewis} and $-0.046 \pm 0.019$ \cite{leinweber}. 

\begin{figure}[htb]
\centerline{\epsfig{file=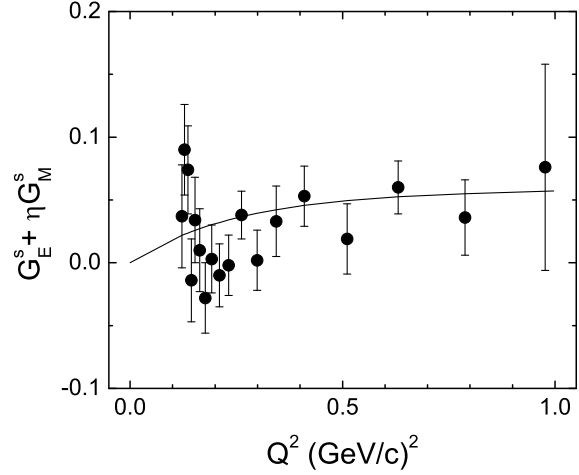,width=0.5\textwidth}} 
\caption[]{\small
Comparison between theoretical and experimental values of 
strange form factors $G_E^s + \eta G_M^s$.  
The experimental values were measured by the G0 Collaboration 
\cite{Armstrong}.}
\label{G0}
\end{figure}

The strange form factors determined in the PVA4 \cite{Maas}, HAPPEX 
\cite{Aniol04,Aniol05b} and G0 \cite{Armstrong} experiments correspond 
to a linear combination of electric and magnetic form factors.   
Fig.~\ref{G0} shows the results for the strange form factor combination 
$G_E^s + \eta G_M^s$ measured recently by the G0 Collaboration at forward angles 
\cite{Armstrong}. A comparison of the calculated values in the two-component model 
with the PVA4 and HAPPEX data shows a similar good agreement as for the G0 data 
shown in Fig.~\ref{G0} \cite{iguazu}. 

\section{Summary and conclusions}

In summary, in this contribution it was shown that the recent experimental 
data on the strange nucleon form factor can be explained very well in 
a two-component model of the nucleon consisting of an intrinsic 
three-quark structure with a spatial extent of $\sim 0.49$ fm 
surrounded by a meson cloud. The present approach is a combination 
of the two-component model of \cite{BI} with the treatment of the 
strange quark content of the vector mesons according to \cite{Jaffe}. 
The parameters in the model are completely determined by the electric and 
magnetic form factors of the proton and neutron. It is noted, that the 
strange couplings do not involve any new parameters. 
On the contrary, the condition that the strange quarks do not contribute to the 
electric charge of the nucleon, leads to an extra constraint relating 
$\beta_{\omega}$ and $\beta_{\phi}$, thus reducing the number of independent 
coefficients of the two-component model of \cite{BI} by one. 

The good overall agreement between the theoretical and experimental 
values for the electromagnetic form factors of the proton and neutron 
and their strange quark content shows that the two-componet model 
provides a simultaneous and consistent description of the electromagnetic 
and weak vector form factors of the nucleon. 

The first results from the SAMPLE, PVA4, HAPPEX and G0 collaborations 
have shown evidence for a nonvanishing strange quark contribution to 
the charge and magnetization distributions of the nucleon. Future 
experiments on parity-violating electron scattering at backward angles 
(PVA4 and G0 \cite{g0back}) and neutrino scattering (FINeSSE \cite{finesse}) 
will make it possible to disentangle the contributions of the different 
quark flavors to the electric, magnetic and axial form factors, and thus 
to provide new insight into the complex internal structure of the nucleon. 

\section*{Acknowledgments}
This work was supported in part by a grant from CONACYT, Mexico.

\end{document}